\documentclass{osa-article}
\usepackage{float}
\journal{osajournal}
\articletype{Research Article}
\usepackage{amsmath}

\begin{document}

\title{Theory of Radiation Pressure on a Diffractive Solar Sail}

\author{Grover A. Swartzlander, Jr.}
\address{Center for Imaging Science, Rochester Institute of Technology, 54 Lomb Memorial Dr., Rochester, NY 14623}
\email{grover.swartzlander@gmail.com}

\begin{abstract}
Solar sails propelled by radiation pressure enable space missions that cannot be achieved using chemical rockets alone.
Significant in-space propulsion for missions such as a solar polar orbiter may be achieved with a sail that 
deviates sunlight at a large average angular direction. 
The momentum transfer efficiency of sunlight diffracted from a sun-facing
diffractive sail comprised of periodic sawtooth prism elements is examined here.
The spectrally averaged efficiency is found to approach that of a monolithic prism
when the grating period is much longer than the peak of the solar spectrum.
This idealized diffraction analysis predicts a greater transverse radiation pressure force 
compared to an idealized reflective sail.  With modern optical design and fabrication
techniques, diffractive solar sails may one day replace reflective sails.
\end{abstract}

\section{Introduction}  \label{Intro}
\noindent
Stellar radiation pressure has scientific origins dating back to 1619 
when Kepler proposed a cause for comet tails \cite{1Schagrin1974,2PlanetarySociety2021}.
Centuries later this phenomenon was found to have the profound effect
of arresting the gravitational collapse of high mass stars 
\cite{3Prialnik2009}.  
Applications to space propulsion were first described in the early 1900's 
by Tsiolkovsky \cite{4Tsiolkovsky1921} and Tsander \cite{5Tsander1924}
whereby a solar sail gains momentum owing to reflected sunlight from a metallic film.
Navigating the heavens by means of solar radiation pressure provides two advantages not
afforded by rocket propulsion: continuous acceleration and an inexhaustible energy source
\cite{6McInnes2004,7Macdonald2006,7Macdonald2014}.
High delta-V missions like solar polar orbiters 
\cite{7Macdonald2006,9Dubill2020,9Swartzlander2021} 
and sub-L1 halo orbits for heliophysics science missions are two examples
\cite{10Johnson2014,11Spencer2020}.
Perhaps counter to one's initial intuition, the component of radiation pressure force
perpendicular to the sun line, rather than the parallel component, is of paramount importance
\cite{6McInnes2004,12Mengali2018,12Swartzlander2017}. 
Whereas the latter component elongates the orbital ellipticity of a sailcraft, the former enables spiral trajectories
that are useful for rendezvous missions to the inner or outer planets. 
For example, navigation toward the sun from a quasi-circular orbit at 1 AU (and at a position 
beyond the influence of Earth gravity) is achieved by tangentially decelerating the sailcraft along its orbit,
resulting in a inward spiral trajectory.
For a sail based on the law of reflection this requires tilting the sail normal away from the sunline,
which has the disadvantage of reducing the solar power projected onto the sail surface.
Nevertheless, solar sails benefit from a seemingly limitless supply of externally supplied photons, unlike
rockets which carry a limited amount of propellant.  Although rockets must be used to loft a solar
sail into space, once there the change of velocity afforded by a sail can greatly exceed that of a rocket
\cite{13Sauer2000,6McInnes2004,14Vulpetti2015}.
This advantage affords opportunities to deliver science instruments to orbits that cannot be reached by rockets alone.
While the technical readiness level of solar sails started to improve in the 1970's 
\cite{15Friedman2015,7Macdonald2014, 10Johnson2014}, 
only recently have demonstration light sail missions been tested in space \cite{16Tsuda2013, 17Johnson2011, 11Spencer2020}.
As of this writing the Planetary Society's Lightsail-2 mission has been circling the earth for three years 
\cite{18Lightsail2}.
If the history of aviation is a guide one may expect many innovations in the design and functionality of solar sails in the coming decades.

Diffractive solar sails were proposed in 2017 as an alternative to reflective sails
 \cite{12Swartzlander2017}.  A potential advantage of a diffractive sail is the
 generation of a significant transverse radiation pressure force while in a sun facing orientation.
 A flat reflective sail must be tilted with respect to the sun line to produce a transverse force.
 The first experimental verification of the transverse force on a diffraction grating
 was reported in 2018 by use of a vacuum torsion oscillator and a laser \cite{19Chu2018} .
 Since then various diffractive sail designs have been explored for both solar
 and laser-driven sails 
 \cite{19Chu2019,19Srivastava2019,19Atwater2019,19Srivastava2020,19Davoyan2021,19Swartzlander2022}.

\section{Radiation Pressure Force}  \label{2}
\noindent
In 1873 Maxwell determined that electromagnetic radiation 
is associated with a pressure that is proportional to the irradiance and
inversely proportional to the speed of light \cite{19Maxwell1873}.  The birth of quantum mechanics
allowed light of wavelength $\lambda$ to be described by particles (photons) 
of momentum $\hbar \vec{k}$ where $\hbar$ is the Planck constant and 
$\vec{k}  = k \hat{k} $ is the wave vector of magnitude $k = 2\pi/\lambda$ 
and unit vector $\hat{k}$ in the direction of propagation.
The radiation pressure force on a light sail may then be described with Newton's third law
which describes conservation of momentum; i.e.,  momentum imparted to a sail
is equal and opposite to the net momentum change experienced by all scattered photons.
For example, if the incident and deviated wave vectors of a photon deviated by the angle $\theta_d$
are respectively expressed $\vec{k}_i = k \hat{z}$ and $\vec{k}_d =$ $k \cos\theta_d \hat{z} + \sin\theta_d \hat{x}$
then the momentum imparted to the sail is given by $\Delta \vec{p}_s =$ $-  \hbar \Delta \vec{k} = $
$\hbar k [ (1 - \cos\theta_d) \hat{z} -  \sin\theta_d \hat{x} ]$.  The sail experiences a positive impulse
along the direction of the sun line,  $\hat{z}$ ranging from $0$ to $2 \hbar k$, and
a transverse component ranging between $\pm \hbar k$.
The irradiance associated with an incident flux of $N$ photons per unit area over a time increment $\Delta t$ 
is given by $I = N \hbar c k / \Delta t$ and thus the force on a sail of area $A$ and projection angle $\psi$ may be expressed
    \begin{equation}
    \begin{split}
    \vec{F}_s =  N A \cos\psi \Delta \vec{p}_s / \Delta t 
    =  (I A \cos\psi /c) \; [ (1 - \cos\theta_d) \hat{z} -  \sin\theta_d \hat{x} ]
    \equiv (I  A/ c) \; \vec{\eta}
    \label{eq:force}
    \end{split}
    \end{equation}
\noindent
where $c$ is the speed of light.
The momentum transfer efficiency vector $\vec{\eta} = \eta_z \hat{z} + \eta_x \hat{x}$
depends on the scattering mechanism  \cite{12Swartzlander2017}.
The component $\eta_x = - \cos\psi \sin\theta_d$ ranges in value from $-1$ to 1, whereas
the component $\eta_z = \cos\psi \; (1 - \cos\theta_d)$ ranges from 0 to 2.  The irradiance
$I$ is the magnitude of the Poynting vector of the illumination source, $\vec{S} = I \hat{z}$,
and $\psi$ is the angle subtending the sunline $\hat{z}$ and the outward surface normal of the sail surface.
In principle the maximum value $|\eta_x| = 1$ may be achieved when the
sail faces the sun $(\psi = 0)$ and simultaneously when light is deviated 
at a right angle to the sun line $(\theta_d = \pm 90^\circ )$.  
Refraction from a prism (see Section \ref{Prism}), for example, may satisfy this condition whereas the law of reflection does not.

\begin{figure}
    \centering
    \includegraphics[scale=0.15]{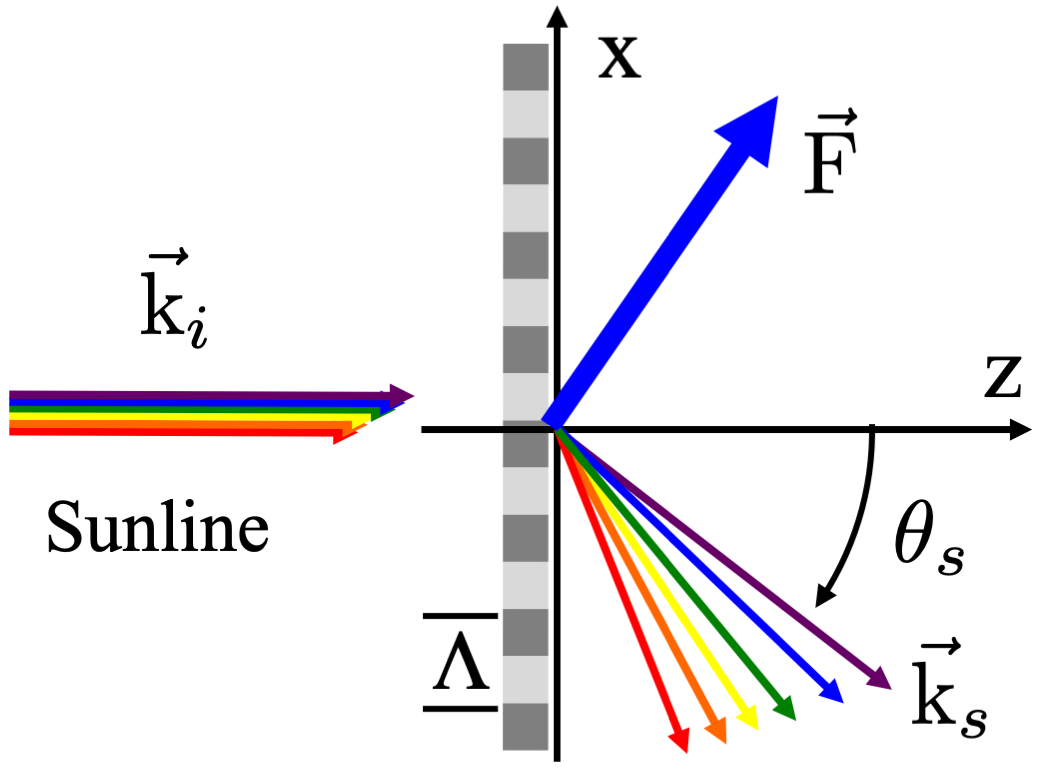}
    \caption{Sunlight incident upon a sun-facing diffractive sail (incidence angle $\psi = 0$) of period $\Lambda$.
    Different wavelengths scatter at different angles $\theta_s$.  
    Each photon of wavelength $\lambda$ is associated with incident wave vector
    $\vec{k}_i$ directed along the sunline $\hat{z}$ and scattered wave vector(s) $\vec{k}_s$.
    The sum of all scattering events produces a radiation pressure force $\vec{F}$.
	}
    	\label{fig:Schematic}
    \end{figure}

In general the direction of scattered light may depend on multiple factors besides incidence angle and wavelength.
However, to gain a baseline understanding of the radiation pressure on a diffraction grating,  
only the normal incidence case is considered below, as illustrated in Fig. \ref{fig:Schematic},
ignoring coherence, polarization, internal and external reflections, diffuse scattering, absorption,
and Doppler shifts.
The net solar radiation pressure force is found by integrating Eq. (\ref{eq:force})
across the solar spectrum, replacing $I$, $\eta_x$, and $\eta_z$ with 
spectral equivalents, $I_\nu$, $\eta_{x,\nu}$, and $\eta_{z,\nu}$,
where $\nu = c / \lambda$ is the optical frequency.

The gravitational attraction of the sun produces an additional force on the sail:
    \begin{equation}
    \vec{F}_g = - ( GM m_s/r^2 ) \; \hat{z}
    \label{eq:gravity}
    \end{equation}
where $m_s$ and $M$ are respectively the mass of the sail and the sun, $G$
is the gravitational constant, and $r$ is the distance between the sun and the sail.
If $F_g >> F_s / c$ then the net force $\vec{F} = \vec{F}_g + \vec{F}_s$ 
has a z-component dominated by gravity and an x-component that is independent of gravity.
This explains why the navigation of a sail typically involves spiral orbits 
\cite{6McInnes2004,14Vulpetti2015,12Mengali2018}
and why here the primary concern is the evaluation of the transverse efficiency $|\eta_x|$.

\section{Refraction from a Right Prism} \label{Prism}
\begin{figure}[t]
    \centering
    \includegraphics[scale=0.12]{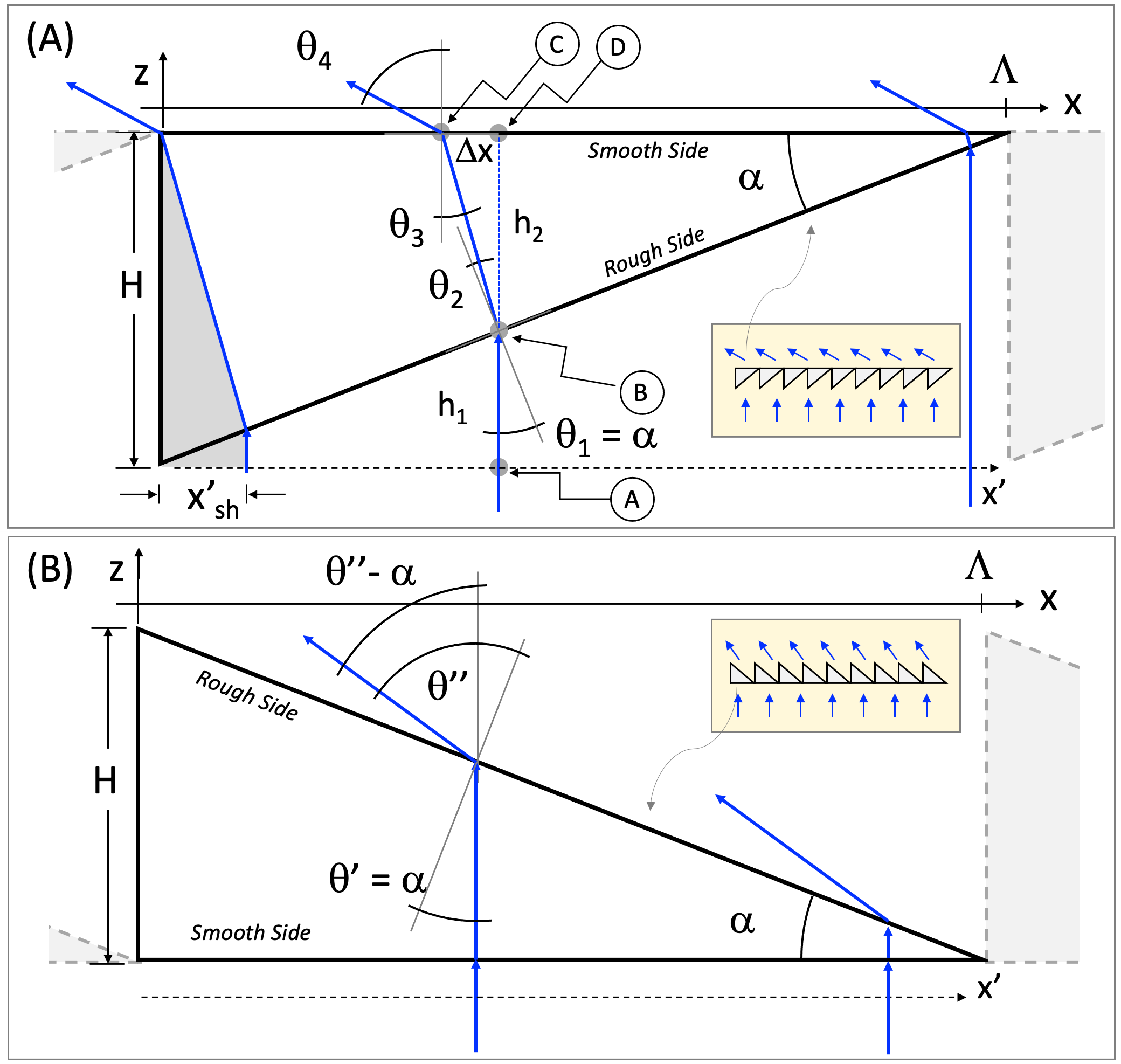}
    \caption{Single element of an array of right prisms 
    of base $\Lambda$, apex angle $\alpha$, and height $H = \Lambda \tan\alpha$
    having refractive index $n$ surrounded by vacuum.  
    Insets depicts periodic gratings.
    (A). Rough side incidence with deviation angle $\theta_d = -\theta_4$.
    First surface angle of incidence: $\theta_1 = \alpha$.  
    Refraction angles $\theta_{2,4}$.
    Second surface angle of incidence $\theta_3 = \alpha - \theta_2$.
    Points A,B,C,D: 
    $(x,z) = ( x+\Delta x, \; - H)$,
    $(x+\Delta x, \; -h_2 )$,
    $(x,0)$,
    $(x+\Delta x, 0)$, where
    $h_2 = H - (x+\Delta x)\tan\alpha$,
    $\Delta x = (H-x\tan\alpha)/(\cot\theta_3+\tan\alpha)$.
    Shadow region width: 
    $x'_{sh} = H/(\cot\theta_3+\tan\alpha)$.
    (B) Smooth side incidence with deviation angle $\theta_d = -(\theta'' - \alpha)$,
     second surface incidence angle $\theta' = \alpha$, and refraction angle $\theta''$.
    }
    \label{fig:Prism}
    \end{figure}

A sawtooth grating may be described as a periodic array of refractive or reflective
right triangles. For the case of a transmission grating described below,
it is instructive to describe the refractive properties of a right prism
having refractive index $n$, base length $\Lambda$, 
height $H$, and prism angle $\alpha = \tan^{-1} (H/\Lambda)$.  
This structure is illustrated in Fig. \ref{fig:Prism} and serves
as the unit cell of a periodic grating in Section \ref{Diffraction2}.
The configuration in Fig. \ref{fig:Prism} (A) is illuminated from the ``rough side'' whereas 
Fig. \ref{fig:Prism} (B) is said to depict ``smooth side'' illumination.  
Both structures are sun-facing $(\psi = 0)$, with the sun line parallel to the $\hat{z}$ direction. 
In both cases light is scattered in the $-\hat{x}$ direction, resulting in a transverse force 
$F_x > 0$ (i.e., $\eta_x > 0$).
For rough-side illumination the incidence and scattering angles are respectively
$\theta_1 = \alpha$ and $\theta_d =  -\theta_4$ where Snell's law and geometry provide
    \begin{subequations} \label{eq:RoughPrism} 
    \begin{align} 
    & \sin\theta_1 = \sin\alpha = n \sin\theta_2  \label{eq:RoughPrisma} \\
    & \theta_3 = \alpha - \theta_2 \label{eq:RoughPrismb} \\
    & n \sin\theta_3 = \sin\theta_4  \label{eq:RoughPrismc} 
    \end{align}  \end{subequations}
The extreme value of $| \theta_d | = 90^\circ$ occurs at the critical condition:
    \begin{subequations}
    \begin{align} 
    \cot \alpha_{cr} = \Lambda_{cr}/H   =  \Lambda / H_{cr} = \sqrt{n^2 -1 \;} -1  \\
    \end{align}  \end{subequations}
where $\alpha_{cr}$ is the prism angle, $\Lambda_{cr}$ is the prism base, and
$H_{cr}$ is the prism height corresponding to the critical angle $| \theta_d | = 90^\circ$.

\begin{figure}
    \centering
    \includegraphics[scale=0.10]{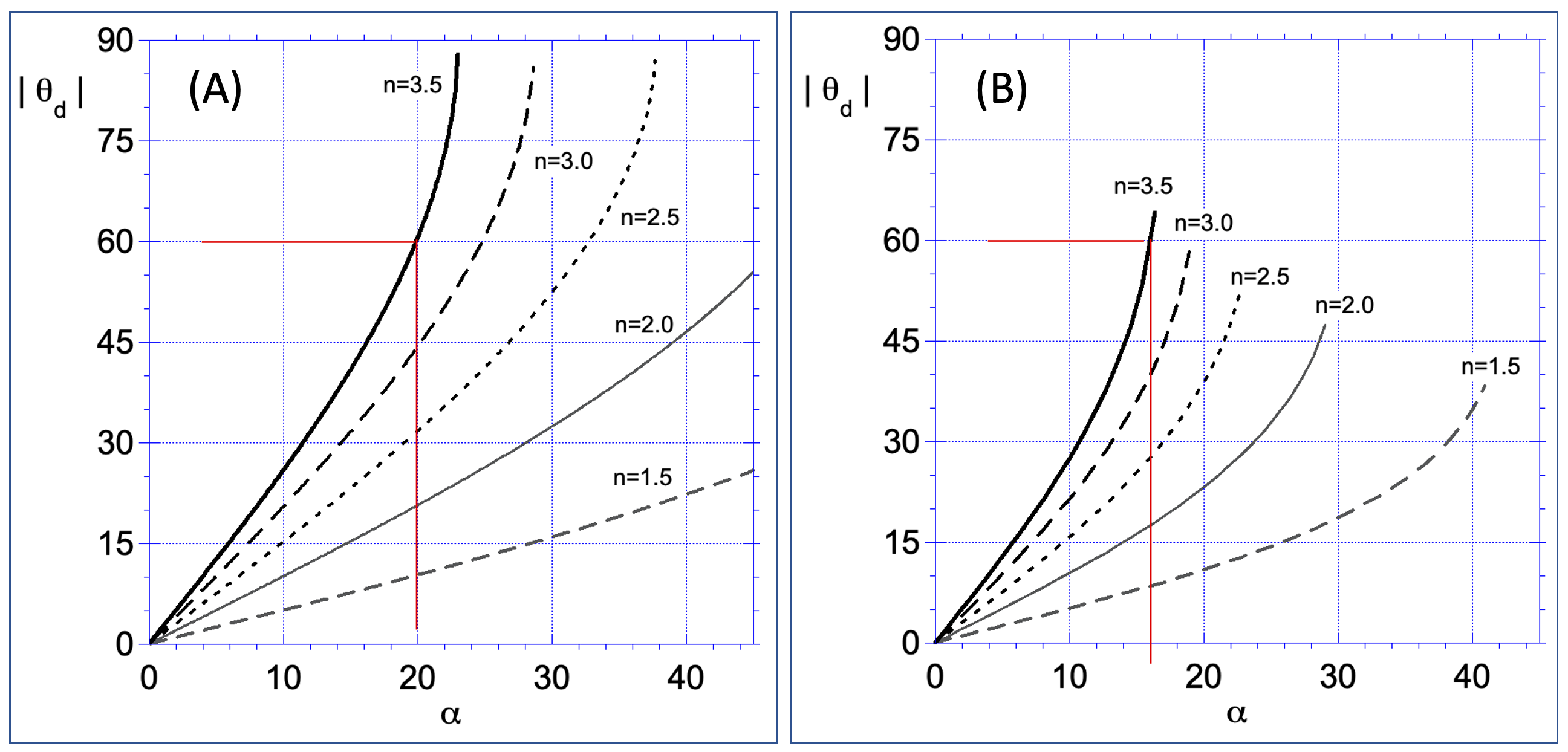}
    \caption{Deviation angle vs.  apex angle for 
    different values of refractive index, $n$.
    (A) Rough sided illumination.  (B) Smooth sided illumination. 
    A deviation of $|\theta_d| = 60^\circ$ (red line) may be possible in both cases, depending on the refractive index.
    }
    \label{fig:Snell2}
    \end{figure}

For smooth-side illumination the transmitted refraction angle $\theta"$ is related
to the prism angle by Snell's law: $n \sin\alpha = \sin\theta"$, and the deviation
angle is given by $\theta_d = -(\theta" - \alpha)$.
Comparisons of the deviation angles for 
rough and smooth sided illumination are plotted in Fig. \ref{fig:Snell2} as
a function of the prism angle $\alpha$ for different values of the refractive index $n$.
Rough-sided illumination clearly produces deviation angles as large as $90^\circ$
whereas the smooth-sided cases do not.  For the latter case the critical angle condition 
$\alpha_{cr} = \sin^{-1} (1/n)$ limits the deviation angle to a maximum value of $90^\circ - \alpha$.
Owing to this limitation only rough-sided illumination is considered below.

For rough-sided illumination a ``shadow region'' of width 
$x'_{sh} = \Lambda \tan\alpha /(\cot\theta_3+\tan\alpha)$ produces rays that are not described
by the discussion above.
The relative extent of $x'_{sh}/\Lambda$ decreases with increasing values of the refractive index.
For example, $x'_{sh}$ is $10 \%$ of $\Lambda$ when $n=3.5$ and $|\theta_d| = 70^\circ$.
Like internally and externally reflected rays, the scattering of rays associated with the
shadow region are ignored below and is left as a topic of future study.
For example, alternative design approaches based on metasurfaces may allow
small surface height modulations that suppress shadowing effects.

If the refractive index varies with optical frequency, the
deviation angle will also vary.  Consequently the net momentum transfer efficiency is
found by integrating across the spectral irradiance distribution $I_\nu$:
    \begin{subequations} \begin{align}
    & \eta_x = - \int_0^\infty I_\nu \; \sin\theta_d(\nu) \; d\nu \Big/ \int_0^\infty I_\nu \; d\nu \\
    & \eta_z =  \int_0^\infty I_\nu \; (1 - \cos\theta_d(\nu) ) \; d\nu \Big/ \int_0^\infty I_\nu \; d\nu 
    \label{eq:GeoDispersion}
    \end{align} \end{subequations}
 where $\theta_d(\nu)$ is determined by use of Eq.s (\ref{eq:RoughPrism}) for each frequency
 and the associated refractive index $n(\nu)$.

\section{Diffraction from a Single Right Prism} \label{Diffraction1}
If the base length is comparable to the wavelength of light
then one must account for diffraction from the single prism element.
At a given optical frequency the transmitted electric field at the output face 
of the prism may be expressed 
    \begin{equation}
    E_\nu (x) = \sqrt{I_\nu \;} \; \exp( i \tilde{k}_x x ) \ \ \text{for } 0 < x < \Lambda
    \label{eq:PrismField}
    \end{equation}
where $\tilde{k}_x = k \sin\theta_d$,
and $I_\nu$ is the incident spectral irradiance measured in units
of $[\text{W/m}^2/\text{Hz}]$, and
$k = 2\pi \nu / c$.
The corresponding far-field (Fraunhofer) distribution may be expressed
    \begin{equation} \begin{split}
    \tilde{E}_\nu(k_x) & = \int_{0}^{\Lambda} \sqrt{I_\nu \;} e^{i\tilde{k}_x x} \; e^{-i k_x x} dx 
    = \Lambda \sqrt{I_\nu \;} e^{i (k_x - \tilde{k}_x) \Lambda/2 } \;
     \frac{\sin ( k_x - \tilde{k}_x ) \Lambda/2 }{( k_x - \tilde{k}_x) \Lambda/2}
    \label{eq:Fraunhofer}
    \end{split} \end{equation}
where $k_x = k \sin \theta$, $\theta$ is the diffraction angle, and $|k_x| \le k$ is required
to afford real values of $k_z = \sqrt{k^2 - k_x^2 \;}$ in the far field $(z >> \pi \Lambda^2 / \lambda)$.  
The field $\tilde{E}_\nu(k_x)$ vanishes if $|k_x| > k$  and it is said to be ``cut off''.

\begin{figure}
    \centering
    \includegraphics[scale=0.4]{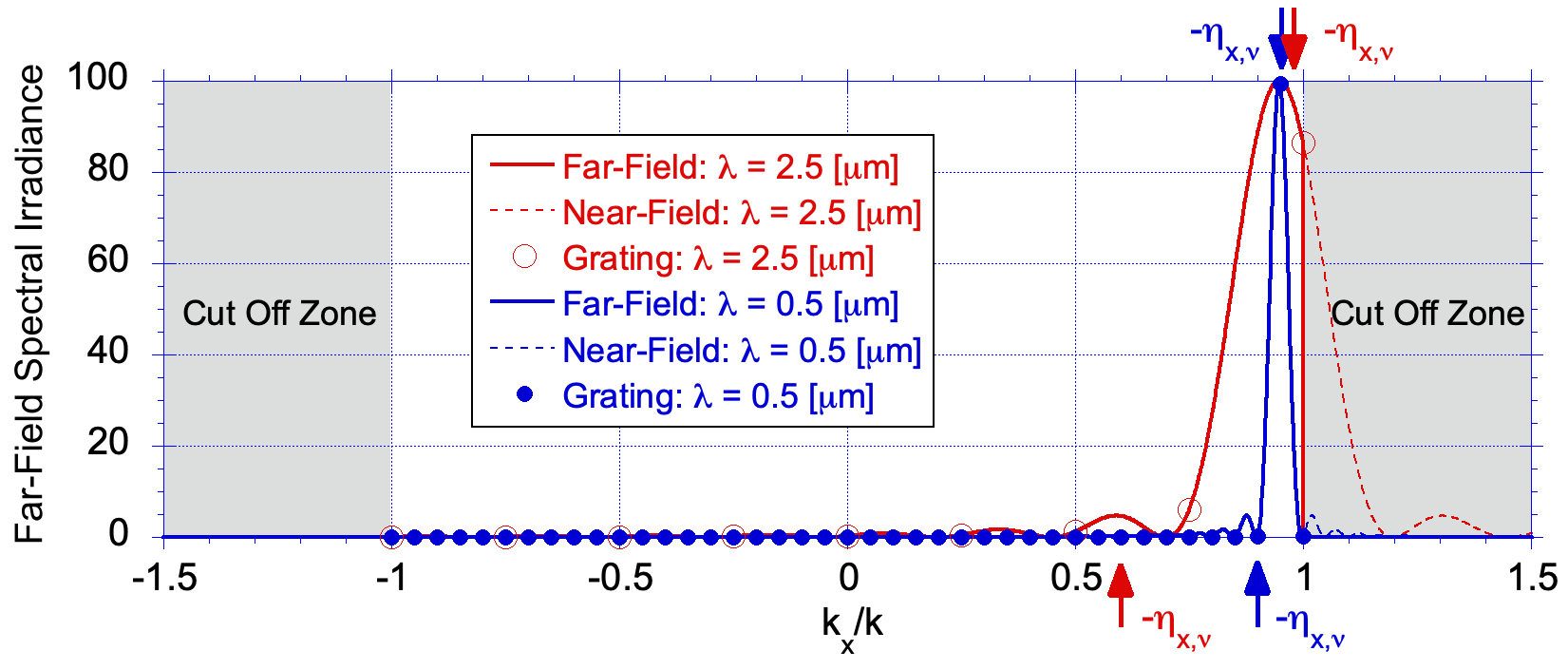}
    \caption{Diffraction from a right prism of base $\Lambda = 10\; [\mu\text{m}]$, 
    height $H = 4\; [\mu\text{m}]$, and refractive index $n = 3.5$ at two wavelengths: 
    $\lambda = 0.5$ (blue) and  $2.5 \; [\mu\text{m}]$ (red). 
    A positive deviation angle is assumed: $\theta_d = 71.4^\circ$.
    Erect arrows: transverse efficiencies $\eta_x$ relative to
    the incident power.  Discrete points correspond to diffraction orders of
    a right triangular grating of period $\Lambda = 10\; [\mu\text{m}]$.
    Inverted arrows: grating transverse efficiencies.
    }
    \label{fig:71deg}
    \end{figure}

Examples of the far-field irradiance $\tilde{I}_\nu(k_x) = | \tilde{E}_\nu (k_x)|^2$
are plotted in Fig. \ref{fig:71deg} at $\lambda = 0.5 \; [\mu\text{m}]$ (blue line)
and at $2.5 \; [\mu\text{m}]$ (red line) for $I_\nu = 1 \; [\text{W/m}^2/\text{Hz}]$.
Nearly half the irradiance distribution is cut off in the long wavelength case.
For this example a deviation angle $\theta_d = 71.4^\circ$ was assumed,
corresponding to $n=3.5$, $\Lambda = 10 \; [\mu\text{m}]$, and $H=4\; [\mu\text{m}]$.
Unlike the above geometric optics descriptions, the deviation is not single-valued
except in the case $\lambda << \Lambda$.
Consequently the net momentum imparted to the sail at a particular value of
frequency must be obtained by integration, resulting in a net spectral efficiency vector with components 
$\eta_{x,\nu} = - < k_{x,\nu} > /k$ and 
$\eta_{z,\nu} = 1- < k_{z,\nu} > /k$
where the effective wave vector components may be expressed
    \begin{subequations}
    \begin{align}
    & < k_{x,\nu} > \; = \;  \int_{-k}^k k_x \; \tilde{I}_\nu(k_x) \; dk_x  
    	\Big/ I_\nu \Lambda
    	\label{eq:etax}
    \\
    & < k_{z,\nu} > \;  = \;  \int_{-k}^k \sqrt{k^2 - k_x^2 \;} \; \tilde{I}_\nu(k_x)  \; dk_x
    	\Big/ I_\nu \Lambda
    	\label{eq:etaz}
    \end{align} 
    \end{subequations}
\noindent
where $\tilde{I}_\nu(k_x) =  |\tilde{E}_\nu(k_x)|^2$ is the far-field spectral irradiance distribution
(see Eq. (\ref{eq:Fraunhofer})), and Parseval's theorem has been used to express the denominators:
    \begin{equation} 
    \int_{-\infty}^\infty | \tilde{E}_\nu(k_x) |^2 dk_x = \int_0^\Lambda |E_\nu (x)|^2 dx = I_\nu \Lambda
    \label{eq:Parseval}
    \end{equation}
Owing to the high spatial frequency cut off,
the far-field centroid $< k_{x,\nu} >$ is shifted to smaller values
of $|k_x|$ compared to the geometrical optics result, 
as illustrated by the erect arrows in Fig. \ref{fig:71deg} for
the case $\theta_d = 71.4^\circ$.
These shifts may be expressed in terms of effective scattering angles, 
$\theta_{\text{eff},\nu} = \sin^{-1} (<k_{x,\nu}> / k)$.
Whereas geometrical optics provides $k_{x}/k = \sin\theta_d = \sin 71.4^\circ$, 
the physical optics cases provide
$<k_{x,\nu}>/k = \sin 63.0^\circ$ at $\lambda = 0.5 \; [\mu\text{m}]$, and
$\sin 37.2^\circ$ at $2.5 \; [\mu\text{m}]$,
i.e., the effective scattering angles are respectively $\theta_{\text{eff},\nu} = 63.0^\circ$ and $37.2^\circ$.

\section{Diffraction from a Right Prism Grating} \label{Diffraction2}
A diffraction grating comprised of an infinite array for right prisms, coherently illuminated
from the rough side as illustrated in the inset of Fig. \ref{fig:Prism}(A), is expect
to produce discrete diffraction orders $k_m$ confined to the envelope function
described by the modulus of Eq. (\ref{eq:Fraunhofer}).  
Similar to the field in Eq. (\ref{eq:PrismField}),
the piece-wise continuous electric field at the transmitting interface may be expressed:
    \begin{equation}
    E_\nu (x) = \sqrt{I_\nu } \; \exp(  i \tilde{k}_x \; (x-x_j) )
     \; , \;
    x_j \le x < x_{j+1} = x_j + \Lambda \;, \text{for} \ \ j=0,\;1,\;2,\; ... \; N-1
    \label{eq:E_j}
    \end{equation}
where 
$x_j = j \Lambda$, $\tilde{k}_x = k \sin\theta_d$,
$k = 2\pi \nu/c$,
the number of periods $N$ tends toward infinity.
The periodicity of the field $E_\nu (x+\Lambda) = E_\nu (x)$ affords the application of 
Fourier series analysis to describe the field beyond the exit face $(z \ge 0)$.
Eliminating non-propagating (evanescent) modes,
the propagating field may be expressed as a finite series of harmonic plane waves:
    \begin{subequations}\label{eq:E}
    \begin{align}
    & {E}_\nu (x,z)  = \sum_{m = M^-_\lambda}^{M^+_\lambda} \tilde{E}_{m,\nu} \; e^{i  m K x } e^{i k_{m,z} z}
    	\label{eq:E_x,z} \\
    & \tilde{E}_{m,\nu}  =  N \int_0 ^ \Lambda E_\nu(x) \; e^{-i  m K x} \; dx 
    	=  A_\nu
    	\frac{\sin(\Delta k_m \Lambda/2)}{\Delta k_m \Lambda/2} 
    	\; e^{-i \Delta k_m \Lambda / 2}
    	\label{eq:Em}
    \end{align}
    \end{subequations}
where $K = 2\pi/\Lambda$, $m$ is an integer, $A_\nu = N \Lambda \sqrt{I_\nu}$,
and
    \begin{equation}
    \Delta k_m = mK - \tilde{k}_x
    	\label{eq:Delta-k-m}
    \end{equation}
The summands in Eq. (\ref{eq:E_x,z}) represent tilted plane wave of amplitude $\tilde{E}_{m,\nu}$ 
and phase $mKx + k_{m,z} z$, with wave vector
    \begin{equation}
    \vec{k}_m = k_{x,m} \; \hat{x} + k_{m,z}\; \hat{z} = k \sin\theta_m \; \hat{x} + k \cos\theta_m \; \hat{z}
    	        = mK \; \hat{x} + k\sqrt{1-(m\lambda/\Lambda)^2} \; \hat{z}
    	\label{eq:k-m}
    \end{equation}
An examination of the $x$-component in Eq. (\ref{eq:k-m}) provides the so-called
grating equation for normal incidence:
$\sin\theta_m = mK/k = m \lambda/\Lambda$. 
Similarly the $z$-component provides
$\cos\theta_m = \sqrt{1 - (m\lambda/\Lambda \;)^2}$.
The diffractive cut off condition corresponds to  
$k_{m,z} = 0$, or equivalently, to mode numbers
    \begin{equation}
    M^\pm_\lambda = \pm \text{INT} [ \Lambda / \lambda ]
    	\label{eq:Mcut}
    \end{equation}
where $\text{INT}$ represents the integral value of the argument, rounded toward zero.
The right hand side of Eq. (\ref{eq:Em}) provides a peak value of $|\tilde{E}_{m,\nu} | = A_\nu$ if
$\Delta k_m = 0$, the latter requiring $(\Lambda/\lambda)\sin\theta_d$ to be integer valued.

The diffraction orders of the propagating field corresponding to an infinite array
of right triangular prisms of period $\Lambda = 10 \; [\mu \text{m}]$ is 
depicted in Fig. \ref{fig:71deg} for $\lambda = 0.5$ and $2.5 \; [\mu \text{m}]$.
The short (long) wavelength case supports mode numbers ranging from $\pm 20$
($\pm 4$).  The irradiance values shown in Fig. \ref{fig:71deg} have been scaled
by $1/N^2$ to aid the eye.  The centroid values for these discrete cases are 
$\eta_{x,\nu} = < k_{x,\nu} >/k = 0.950$  for $\lambda = 0.5 \; [\mu \text{m}]$ and
$\eta_{x,\nu} = < k_{x,\nu} >/k = 0.982$ for $\lambda = 2.5 \; [\mu \text{m}]$,
respectively corresponding to effective deviation angles 
$\theta_\text{eff} = 71.8^\circ$ and $79.1^\circ$.  
These values are greater than those in the single-grating cases described at the end of Section \ref{Diffraction1}
because little light has been cut off.

\begin{figure}[b]
    \centering
    \includegraphics[scale=0.18]{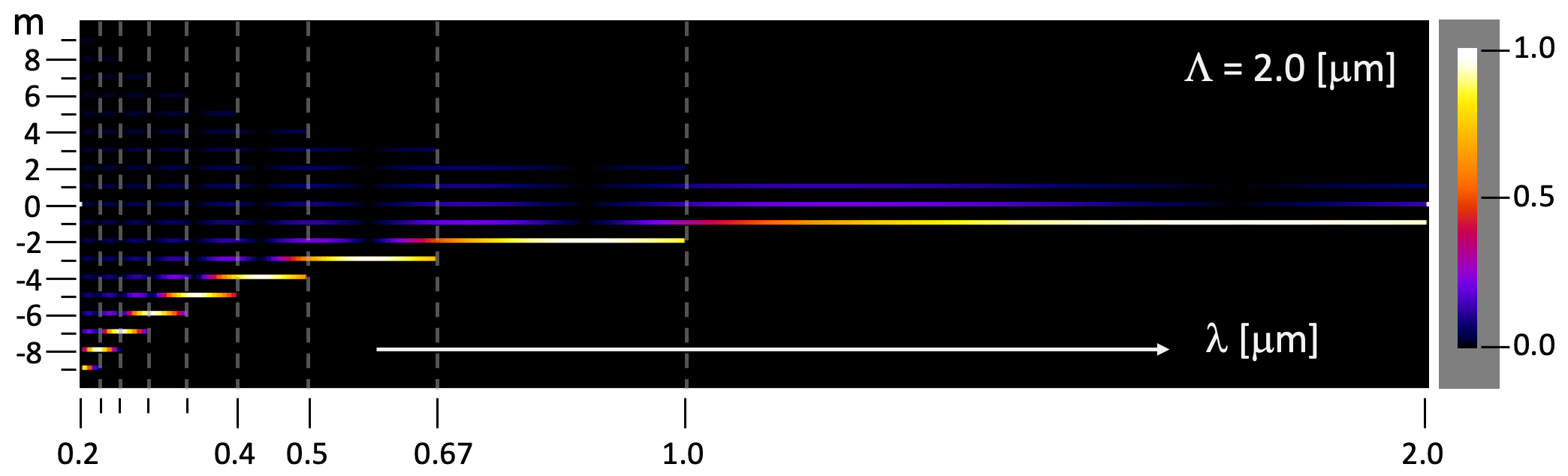}
    \caption{Spectral diffraction amplitudes $|\tilde{E}_{m,\nu}/A_\nu |$ for 
    a right prism grating of period $\Lambda = 2 \; [\mu\text{m}]$
    and refractive deviation angle $\theta_d = -60^\circ$.
    Maxima occur at $\lambda = (\Lambda/m)\sin\theta_d$.
    Cut off modes are not shown.
    }
    \label{fig:SpectralDiffractionModes}
    \end{figure}

Solutions of Eq. (\ref{eq:Em}) are depicted in Fig. \ref{fig:SpectralDiffractionModes} for the wavelength
range $\lambda: (0.2 \; [\mu\text{m}] , \; 2.0 \; [\mu\text{m}])$, 
without the inclusion of the cut-off modes,
for the case $\Lambda = 2 \; [\mu\text{m}]$ and  $\theta_d = -60^\circ$.
Salient features for this case, such as the wavelength of peak diffraction orders and cut-off mode numbers
are listed in Table \ref{table:Table1}.
As expected for a negative deviation angle, the dominate diffraction orders correspond to negative values of $m$.

The diffraction spectrum exhibits cut-off wavelengths at the wavelength values 
$\Lambda$, $\Lambda/2$, $\Lambda/3 \;...$,
or equivalently at the frequencies $\nu_0$, $2 \nu_0$, $3 \nu_0, \;...$,
where $\nu_0 = c/ \Lambda$ is the fundamental frequency associated with the grating. 
These boundaries are marked with vertical dashed lines in Fig. \ref{fig:SpectralDiffractionModes}
and are listed as $\lambda''_{m'}$ in Table \ref{table:Table1}.
The wavelength at which $| \tilde{E}_{m'} |$ is maximum, i.e., where
$| \tilde{E}_{m'} |  = A_\nu \; \delta_{m,m'}$ (where $\delta_{m,m'}$ is the Kronecker delta function),
is found by setting $\Delta k_{m'} \Lambda/2 = 0$, i.e,
at $\lambda_\text{peak} = (1/m') \Lambda \sin\theta_d$ (see Eq. (\ref{eq:Delta-k-m})).
The range between 
$\lambda'_{m'} = \Lambda \sin\theta_d / (m' - 1)$ and $\lambda''_{m'} = -\Lambda/m' $
shall be called the $m'^{ \; \text{th}}$ dominant band.
For example the $m' = -2$ dominant band ranges from 
$\lambda'_{m'} = 0.577 \; [\mu\text{m}]$ to 
$\lambda''_{m'} = 1.0 \; [\mu\text{m}]$.
It is evident in Table \ref{table:Table1} that as the magnitude of the dominant mode number $|m'|$
increases, the magnitude of $|\theta_{m'}|$ at $\lambda'_{m'}$ increases.
For example $\theta_{m'=-2} = -35.3^\circ$ whereas $\theta_{m'=-8} = -50.3^\circ$.
This suggests that on average, short wavelength light diffracts at higher angles, 
approaching the refraction angle $\theta_d$, 
compared to long wavelength light (as expected from the geometric optics approximation
which assumes $\lambda \rightarrow 0$.)
The importance of this for solar sailing will become evident in the discussion below.
    \begin{table}
    \centering
    \begin{tabular} 
    { | p{0.6 cm} || p{1.5cm} | p{2.2cm} | p{2.1cm} | p{0.8cm} | p{2.2cm} | }
     \hline
    $ \ \ \ m'$ &   $ \  \lambda_{\text{peak}} \; [\mu \text{m}]$ &
    $(\lambda'_{m'} , \; \lambda''_{m'}) \; [\mu \text{m}]$ & $\ \ \ \ \ \ \     -\theta_{m'} $ & 
     	$ \ M^\pm_\lambda $ & $\ \sin^{-1}(\frac{M^\pm \lambda_{\text{peak}}}{\Lambda})$
    	\\
     \hline
    $ \  -1 $  &   $ \  1.732 $ &     $ \ \ 0.866 , 2.000 $ &  $ \ 25.7^\circ , 60^\circ, 90^\circ $ & $ \ \pm 1$    & $ \ \pm 60.0^\circ$  \\
    $ \  -2 $  &   $ \  0.866 $ &     $ \ \ 0.577 , 1.000 $ &  $ \ 35.3^\circ , 60^\circ, 90^\circ $ & $ \ \pm 2$    & $ \ \pm 60.0^\circ$   \\
    $ \  -3 $  &   $ \  0.577 $ &     $ \ \ 0.433 , 0.667 $ &  $ \ 40.5^\circ , 60^\circ, 90^\circ $ & $ \ \pm 3$    & $ \ \pm 60.0^\circ$  \\
    $ \  -4 $  &   $ \  0.433 $ &     $ \ \ 0.346 , 0.500 $ &  $ \ 43.9^\circ , 60^\circ, 90^\circ $ & $ \ \pm 4$    & $ \ \pm 60.0^\circ$  \\
    $ \  -5 $  &   $ \  0.346 $ &     $ \ \ 0.289 , 0.400 $ &  $ \ 46.2^\circ , 60^\circ, 90^\circ $ & $ \ \pm 5$    & $ \ \pm 60.0^\circ$  \\
    $ \  -6 $  &   $ \  0.289 $ &     $ \ \ 0.247 , 0.333 $ &  $ \ 47.9^\circ , 60^\circ, 90^\circ $ & $ \ \pm 6$    & $ \ \pm 60.0^\circ$  \\
    $ \  -7 $  &   $ \  0.247 $ &     $ \ \ 0.217 , 0.286 $ &  $ \ 49.3^\circ , 60^\circ, 90^\circ $ & $ \ \pm 8$    & $ \ \pm 81.8^\circ$  \\
    $ \  -8 $  &   $ \  0.217 $  &    $ \ \ 0.192 , 0.250 $ &  $ \ 50.3^\circ , 60^\circ, 90^\circ $ & $ \ \pm 9$   & $ \ \pm 77.0^\circ$  \\  
     \hline
    \end{tabular}
    \caption{
    Dominant band, $m'$ with corresponding peak wavelengths $\lambda_{\text{peak}}$ where $|\tilde{E}_m'|/A_\lambda = 1$.
    Wavelength range of the dominant mode $(\lambda'_{m'} , \; \lambda''_{m'})$.
    Diffraction angles $\theta_{m'} = \sin^{-1}(m'\lambda/\Lambda)$ for $\lambda: \; \lambda'_{m'}, \; \lambda_{\text{peak}}$,
    and $\lambda''_{m'}$.
    Grating period $\Lambda = 2 \; [\mu\text{m}]$.
    Last two columns: Cut-off mode values $M^\pm_\lambda$ and corresponding diffraction angles at $\lambda = \lambda_{\text{peak}}$.
    }
    \label{table:Table1}
    \end{table}

\section{Momentum Transfer Efficiency of a Sunlit Diffractive Sail} \label{MTE}
\noindent
The net force and efficiency 
may be determined by integrating over all scattered light.
Here the illumination is represented by the solar black body spectrum
with plane waves normally incident upon the rough side of the grating.
The small angular extent of the the sun (e.g., 
$\Delta\theta_{sun} = 0.5^\circ$ at 1 AU) is ignored.
The spatial coherence 
length $L_c = 2 \lambda / \pi \Delta\theta_{sun}$
is also assumed to be large enough compared to the grating period
to afford narrowing of the diffraction peaks, e.g., 
$L_c/\Lambda > 4$.  This condition is satisfied for $\Lambda = 10 \;[\mu\text{m}]$ 
if $\lambda > 0.55 ;[\mu\text{m}]$ (i.e, $\nu = c/\lambda < 550$ THz) 
which includes more than half the power emitted by the sun.
In comparison, a grating period of $\Lambda = 2 \;[\mu\text{m}]$ requires 
$\lambda > 0.11 ;[\mu\text{m}]$ (i.e, $\nu < 2740$ THz) which includes nearly 
the entire solar black body spectrum.
Hence, for grating periods $\Lambda \le 10 \;[\mu\text{m}]$, the results
in Sec. \ref{Diffraction2} will be used to determine the radiation pressure force.
A full coherence analysis is beyond the scope of this report.

Numerical integration across all diffraction orders and a 
broad band of wavelengths is required to determine the net 
momentum transfer efficiency.
As illustrated in Fig. \ref{fig:SpectralDiffractionModes}
the interval between cut off frequencies is not regular.
On the other hand, the interval between corresponding
frequencies is regular, with cut off frequencies separated by
$\nu_0 = c/\Lambda$.
Numerical integration across optical frequencies is therefore made 
to insure regular sampling $\delta \nu = \nu_0 / N_\nu$
when employing the trapezoid rule, where $N_\nu > 100$ samples was used.

The black body spectral irradiance
distribution in frequency space differs from that in wavelength space, providing
a peak value (Wien's displacement law) at $\nu_\text{peak} = (5.879 \times 10^{10} \; [\text{Hz/K}]) \;  T$.
At the solar temperature $T = 5770 \; [\text{K}]$, $\nu_\text{peak} = 339.2 \; [\text{THz}]$ 
(which corresponds to the wavelength $c/ \nu_\text{peak} = 884.4 \; [\text{nm}]$).
The black body spectral irradiance a distance $r$ from the sun may be expressed
in units of $[\text{W/m}^2/\text{Hz}]$:
    \begin{equation}
    I_\nu = \frac{R_S^2}{r^2} \; \frac{2 \pi h \nu^3}{c^2} \; \frac{1}{\exp(h \nu / k_B T) - 1}
    	\label{eq:blackbody-nu}
    \end{equation}
where $R_S = 6.957 \times 10^{8} \; [\text{m}]$ is the solar radius,
$h = 6.626 \times 10^{-34} \; [\text{J} \cdot \text{s}] $ is the Planck constant, and 
$k_B = 1.381 \times 10^{-23} \; [\text{J/K}]$ is the Boltzmann constant.
Numerically integrating Eq. (\ref{eq:blackbody-nu}) across all wavelengths provides 
the so-called solar constant at the orbital radius of the earth ($r_E = 1.496 \times 10^{11} \;[\text{m}]$):
$I_E = 1.36 \; [\text{kW/m}^2]$.   This result is validated by the value expected from the Stefan-Boltzmann
law: $I_E = (R_s / r_E)^2 \sigma T^4$, where $\sigma = 5.670 \times 10^{-8} \; [\text{W/m}^2/\text{K}^4]$
is the Stefan-Boltzmann constant.
A rectangular sail of length $L_x$ and width $L_y$ is assumed, with diffractive scattering in the 
$\pm x-$direction as described above.
For convenience the net radiation pressure force at $r=r_E$ is described below.

Owing to orthogonality between diffraction orders, the net force $\vec{F}$
at $1 \; [\text{AU}]$ from the sun
may be decomposed into force components $\vec{F}_{m,\nu}$
attributed to each $m^\text{th}$ order diffraction mode
for a given frequency $\nu$:
    \begin{subequations}\label{eq:Frp0}
    \begin{align}
    & \vec{F}= \int_0^\infty \vec{F}_\nu \; d\nu 
    	 = \int_0^\infty \sum_{m=M^-_\nu}^{M^+_\nu} \vec{F}_{m,\nu} \; d\nu
    	\label{eq:Frpa} \\
    & \vec{F}_{m,\nu} = I_{m,\nu} \; \frac{\vec{k}_{\text{i}} - \vec{k}_{m}}{ck} \; L_x L_y
    	\;, \ \ m:[M^-_\nu , M^+_\nu]
    	\label{eq:Frpb} 
    \end{align}
    \end{subequations}
where $I_{m,\nu} = |E_{m,\nu}|^2$ is the spectral irradiance of the $m^\text{th}$ mode
described by Eq.s (\ref{eq:Em}, \ref{eq:blackbody-nu}),
$\vec{k}_{\text{i}} = k \hat{z}$ is the incident wave vector, 
$\hat{z}$ is a unit vector directed radially from the sun (see Fig. \ref{fig:Schematic}), and
$\vec{k}_{m}$ 
is the scattered wave vector of the $m^\text{th}$ mode described in Eq. (\ref{eq:k-m}).
The cutoff modes $M^\pm_\nu$ are given in Eq. (\ref{eq:Mcut}) with $\lambda$ replaced with $c/\nu$.

The momentum transfer efficiency
is defined as the ratio of the net force, Eq. (\ref{eq:Frpa}),
and the normalizing force parameter $P_{\text{in}}/c$, where $P_{\text{in}}$ is the net power projected on a 
sun-facing sail:
    \begin{subequations}\label{eq:eta0}
    \begin{align}
    & \vec{\eta} = \frac{\vec{F}} {P_{\text{in}}/c}
    	 = \int_0^\infty \sum_{m=M^-_\nu}^{M^+_\nu} \vec{\eta}_{m,\nu}  \; d\nu
    	 = \int_0^\infty \; \vec{\eta}_\nu \; d\nu
    	\label{eq:eta} \\
    & \vec{\eta}_{m,\nu} = \frac{\vec{F}_{m,\nu} }{ P_{\text{in}} /c} =
    	\frac{I_{m,\nu}}{I_E}  \left[ (1-\sqrt{1-(m\lambda/\Lambda)^2} \; )\hat{z} + (m \lambda/\Lambda) \hat{x} \right]
    	\;, \ \ m:[M^-_\nu , M^+_\nu]	
    	\label{eq:eta-m-lambda}  \\
    & \vec{\eta}_\nu = \sum_{m=M^-_\nu}^{M^+_\nu} \vec{\eta}_{m,\nu} 
    	\label{eq:eta-lambda} 
    \end{align}
    \end{subequations}
where $ \vec{\eta}_{m,\nu}$  is the modal spectral efficiency,
$\vec{\eta}_\nu$ is the spectral efficiency, and where
    \begin{subequations}\label{eq:irradiance}
    \begin{align}
    & I_{m,\nu} = |E_{m,\nu}|^2 = I_\nu \; \frac{\sin^2(\Delta k_m \Lambda / 2)} {(\Delta k_m \Lambda / 2)^2}
    	=  I_\nu \; \text{sinc}^2 (  \pi m -  \pi \sin\theta_d \; \Lambda / \lambda  )
    	\label{eq:I-m-lambda} \\
    & {I}_E = P_{\text{in}}/ L_x L_y = \; \int_0^\infty I_{\nu} \, d\nu  = 
    	\; \sum_{m=-\infty}^{\infty} \int_0^\infty I_{m,\nu} \, d\nu = 1.36 \; [\text{kW/m}^2]
    	\label{eq:solar-constant} 
    \end{align}
    \end{subequations}
Using Eq.s (\ref{eq:I-m-lambda}, \ref{eq:blackbody-nu}) and $\lambda = c/\nu$ 
the values in Eq. (\ref{eq:eta0})
may be numerically integrated.
    \begin{figure}
    \centering
    \includegraphics[scale=0.20]{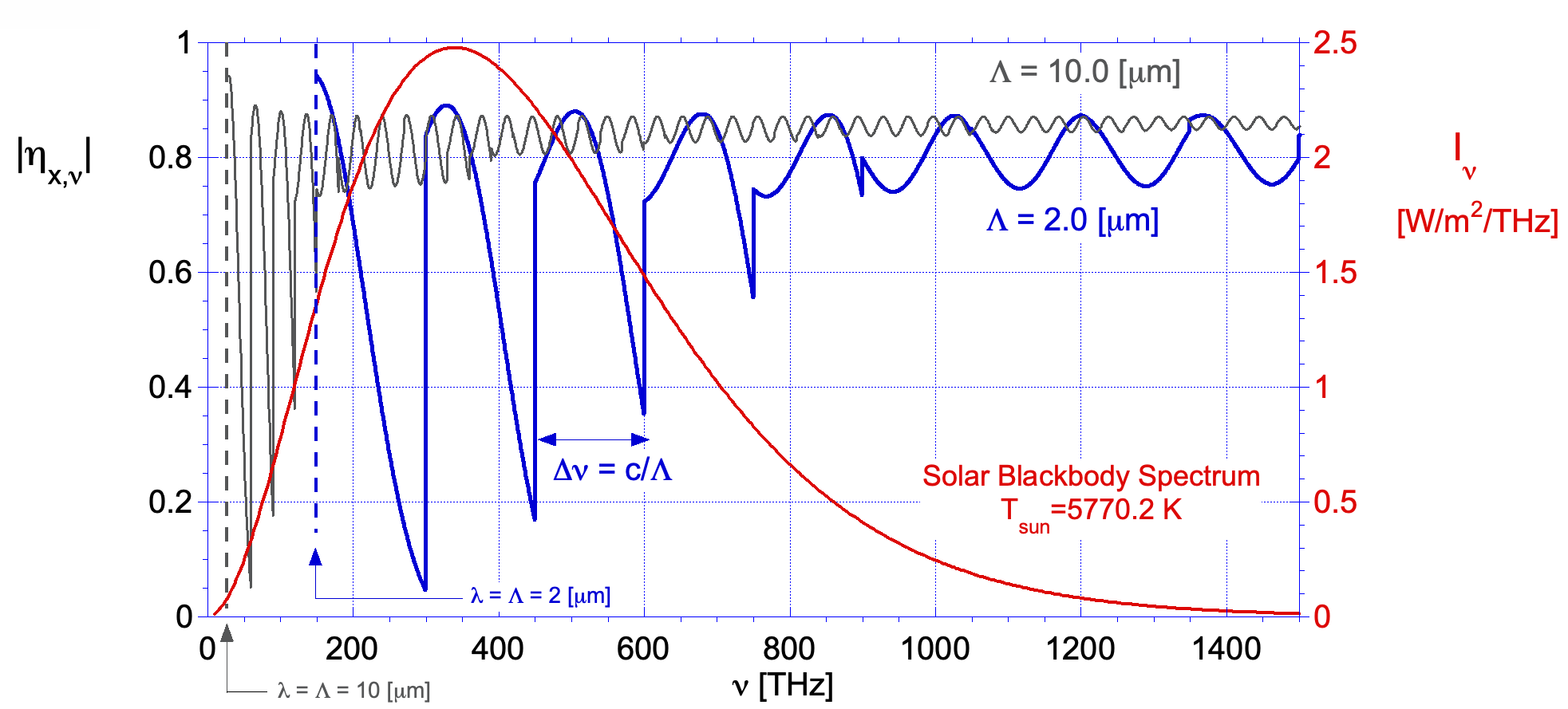}
    \includegraphics[scale=0.20]{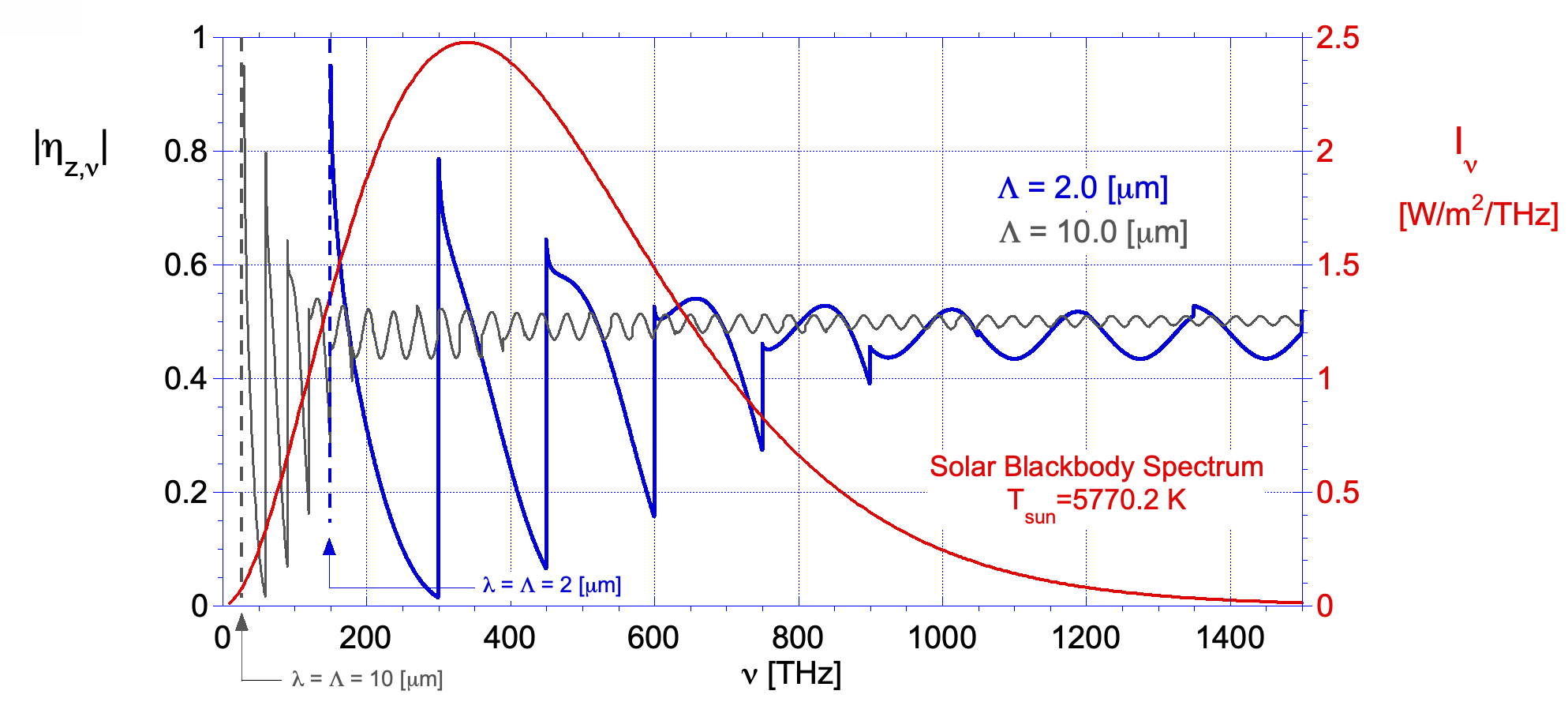}
    \caption{Spectral momentum transfer efficiency components, 
    $\eta_{x,\nu}$ and $\eta_{z,\nu}$ for grating periods
    $\Lambda = 2$ and $10 \; [\mu\text{m}]$ and refractive deviation
    angle $\theta_d = -60.5^\circ$.
    Cut-off wavelengths are separated by equal frequency intervals $\Delta\nu = c/\Lambda$.
    }
    \label{fig:SolarGratingEfficiency-eta,x,z}
    \end{figure}
    
The momentum transfer efficiency for two cases, $\Lambda = 2 \; \text{and} \; 10 \; [\mu\text{m}]$ are
depicted in Fig. \ref{fig:SolarGratingEfficiency-eta,x,z} for a grating of index $n=3.5$,
apex angle $\alpha= 20^\circ$, and refractive deviation angle $\theta_d = \pm 60.5^\circ$.
The value $\nu_0 = c/\Lambda$ also corresponds to the low frequency (long wavelength) cut-off,
below which $\eta_{x,\nu} = 0$.
At frequencies great than, but close to the cut off (with corresponding wavelength 
less than but on the same order as the grating period),
strong diffraction effects are evidenced by pronounced variations of the efficiency values.
The transverse spectral efficiency $\eta_{x,\nu}$  varies with frequency, reaching an extremum 
of roughly $ |{\eta}_{x,\nu} | = 0.94$ at $\nu = \Delta\nu$, 
falling to almost zero at $\nu = 2 \Delta\nu$ in Fig. \ref{fig:SolarGratingEfficiency-eta,x,z}(A).
On the other hand, for large frequency values (wavelengths much smaller than the grating period), 
diffractive modulations are less pronounced,
with the transverse efficiency converging toward the value for pure refraction:
 $| {\eta}_{x,\nu} | \approx | \sin\theta_d | = 0.87$.
As suggested in the discussion of Table \ref{table:Table1}, the average magnitude of efficiency indeed
increases with increasing frequency (decreasing wavelength).
Comparing the cut off frequency for the two cases in Fig. \ref{fig:SolarGratingEfficiency-eta,x,z}(A)
it is evident that the longer period grating provides fuller coverage of the solar spectrum,
thereby providing a stronger radiation pressure force.
Similar phenomenon are seen for $\eta_{z,\nu}$ in  Fig. \ref{fig:SolarGratingEfficiency-eta,x,z}(B),
with the magnitude tending toward $\cos\theta_d = 0.49$ at high frequencies.
The reader is reminded that these calculations to not account for reflected light, such as
the backscatter of light in the cut off frequency band.
    \begin{figure}
    \centering
    \includegraphics[scale=0.22]{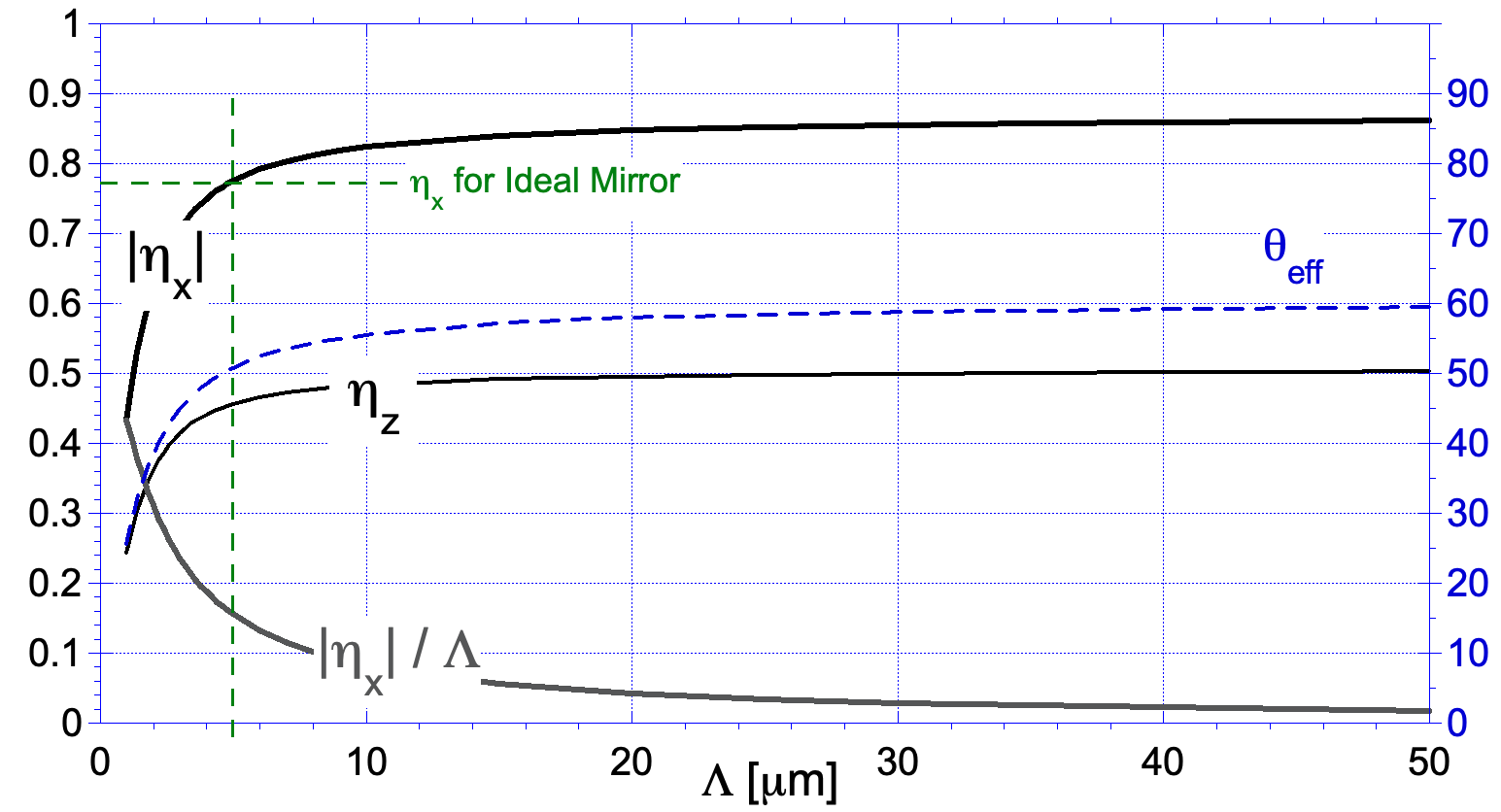}
    \caption{Momentum transfer efficiency components $| \eta_x| $ and $\eta_z$ of a transmission
    grating of apex angle $\alpha = 20^\circ$, refractive index $n=3.5$, and refractive deviation angle
    $|\theta_d | = 60.5^\circ$ for solar black body illumination.  The relation $| \eta_x| = \sin\theta_\text{eff}$
    defines the effective deviation angle $\theta_\text{eff}$ (blue line).  For comparison an ideal mirror
    provides  $|\eta_x| = 0.77$ (green line). Ratio $|\eta_x|/\Lambda$ (gray line).
    }
    \label{fig:eff-vs-period}
    \end{figure}

Integrating the calculated values of spectral efficiency across the optical frequency band 
provides the net efficiency components, $\eta_x$ and $\eta_z$.
Examples of these components are plotted as a function of the grating period $\Lambda$
in Fig. \ref{fig:eff-vs-period}
for the case $n=3.5$, $\alpha = 20^\circ$, $\theta_d = \pm 60.5^\circ$, and solar
black body illumination across the wavelength band $0.1 \; [\mu\text{m}]$ to the cutoff wavelength $\Lambda$.
The magnitude of the transverse efficiency increases with the grating period, with an inflection point at 
$\Lambda \approx 5 \; [\mu\text{m}]$ and an asymptotic value $| \eta_x | = | \sin\theta_d | = 0.87$.
The effective deflection angle of the sunlight may be defined by the relation
$\eta_x = - \sin\theta_{\text{eff}}$.  Values of $\theta_{\text{eff}}$ are plotted in Fig. \ref{fig:eff-vs-period},
illustrating the long period asymptotic values approaching the refractive deflection angle.

The above values may be compared with the momentum transfer efficiency
of an ideal reflective sail:
$\vec{\eta} = 2 \cos^2 \theta_i ( \cos\theta_i \hat{z} - \sin\theta_i \hat{x})$
\cite{6McInnes2004}. At the optimum incidence angle $\theta_i = \pm 35.3^\circ$,
$|\eta_x| = 0.77$, thereby providing an effective deflection angle $|\theta_{\text{eff}} | = 50.3^\circ$.
The foregoing grating analysis based on a series of prisms, 
each having a refractive deflection angle of $\theta_d = \pm 60.5^\circ$,
demonstrates $|\eta_x| > 0.77$ for grating periods $\Lambda > 4.8 \; [\mu\text{m}]$.
Designing the prisms to achieve larger values of $|\theta_d|$ are also expected
support high values of $|\eta_x|$.

For the longitudinal efficiency along the sun line 
$\eta_z = 1.09$ for an ideal mirror, which is significantly greater than the
corresponding value for the analyzed prism grating. 
For orbit changing maneuvers a large value of $\eta_z$ increases the orbital eccentricity, 
which is typically not desirable.
However, as stated in Sec. \ref{Intro} the component of radiation pressure force along the sun line 
is typically negligible compared to solar gravity, and thus the value of $\eta_z$ is not a concern.
Note that a fuller grating analysis that includes reflected light from the grating is expected
to increase the value of $\eta_z$ above those shown in Fig. \ref{fig:eff-vs-period}.

\section{Acceleration} \label{Acceleration}
The foregoing analysis suggests that a long period grating has a greater transverse momentum transfer efficiency,
and therefore a greater force, compared to a short period grating.  
The paramount concern for orbit changing maneuvers in space, however, is the transverse acceleration,
    \begin{equation}
    a_x 
    = \frac{ P_{\text{in}}}{c M_{sc}} \eta_x
    = \frac{I_E A}{c(m' + m_s)} \; \eta_x
    \end{equation}
where $M_{\text{sc}} = m' + m_s$ is the total mass of the spacecraft, and
$m_s$ and $m'$ are respectively the mass of the sail and everything else.  
For a comparatively massive payload, $m' >> m_s$,  the acceleration is independent of the prism dimensions,
but dependent on the sail area:
$a_x \approx (I_E A / m' c) \; \eta_x$.
In contrast, if $m' << m_s$, the acceleration is independent of the sail area, and 
increases rapidly with decreasing prism height $H = \Lambda\tan\alpha$:
$a_x \approx (2 I_E / c \rho_s ) \; \eta_x / \Lambda \tan\alpha$, where
$\rho_s$ is the mass density of the sail material. 

Setting the mass ratio $\mu = m'/m_s$ as a dimensionless design parameter
the acceleration may be expressed:
    \begin{equation}
    a_x =  \frac{2 I_E}{c \rho_s (1+\mu)} \frac{\eta_x}{\Lambda \tan\alpha}
    \end{equation}
The value of $|\eta_x| / \Lambda$, plotted in Fig. \ref{fig:SolarGratingEfficiency-eta,x,z},
indicates that the loss of efficiency at small values of $\Lambda$ is surpassed by the
small mass advantage.  That is, it is sensible to sacrifice efficiency to achieve
a greater acceleration owing to smaller mass.
For example, if $\rho_s = 1000 [\text{kg/m}^3]$, $\alpha = 20^\circ$,
$I_E = 1.36 \; [\text{kW/m}^2]$, $\Lambda = 2 \; [\mu\text{m}]$,  and $\mu =1$,
the numerically calculated transverse efficiency is $\eta_x = 0.62$ and thus
$a_x = 3.9 [\text{mm/s}^2]$, which could provide inclination angle changes 
of roughly $8.5^\circ$ per day for a solar polar orbiter mission \cite{13Sauer2000}.

Comparing $(\eta_x/\bar{H})_\text{diff}$ to $(\eta_x/H)_\text{refl}$, where 
$\bar{H}_\text{diff} = (1/2) \Lambda \tan\alpha$ is the average thickness of the grating,
and $H_\text{refl}$ is the thickness of a reflecting film, the condition
$(\eta_x/\bar{H})_\text{diff} >  (\eta_x/H)_\text{refl}$ suggests greater acceleration
for a diffraction grating if $\bar{H}_\text{diff} / H_\text{refl} <   (\eta_x)_\text{diff} / (\eta_x)_\text{refl}$.
For the ideal case, if $(\eta_x)_\text{refl} = 0.77$ and $(\eta_x)_\text{diff} = 1.0$, this condition
provides $\bar{H}_\text{diff} / H_\text{refl} < 1.3$, thereby requiring a prism height 
$\Lambda \tan\alpha < 2.6 H_\text{refl}$.  For example, if $H_\text{refl} = 3 \; [\mu\text{m}]$
then $\Lambda \tan\alpha < 7.8 \; [\mu\text{m}]$.
What is more, if this condition is expressed
$\Lambda \tan\alpha / H_\text{refl} <   2 (\eta_x)_\text{diff} / (\eta_x)_\text{refl}$,
then equality between the prism height and film thickness 
$(\Lambda \tan\alpha  = H_\text{refl})$ suggests that $(\eta_x)_\text{diff}$ 
may be as small as $(1/2) (\eta_x)_\text{refl}$.
This analysis assumes a massless substrate supporting the array of prisms.
Nevertheless there is a clear acceleration benefit of a lower mass prism grating
having the same height as the thickness of a reflecting film.

\section{Conclusion}  \label{Conclusion}
To navigate within the neighborhood of a few AU's from the sun
via spiral trajectories, a spacecraft must experience thrust perpendicular
to the rays of the sun.  Such a force may be created by means
of radiation pressure whereupon sunlight is deviated by an angle 
approaching $90^\circ$.  Using the law of reflection to deflect
light is counterproductive since the fraction of solar power
projected onto the sail decreases with tilt angle.  In contrast
a diffraction grating may achieve large deviation angles in
a sun-facing orientation.  To demonstrate this the radiation pressure on
an idealized transmission grating comprised of right prisms
has determined using Fourier series analysis for wavelengths
spanning the solar spectrum.
A similar analysis may be made for a reflection grating.
This paper serves as a baseline study with numerous simplifying assumptions.
The primary outcomes are: (1) at wavelengths much smaller
than the grating period (where geometric optics is valid)
the light deviation angle approaches that predicted by
Snell's law; (2) as the wavelength approaches the grating 
period the deviation angle exhibits pronounced wavelength-dependent
modulation and consequently a smaller deviation angle; 
(3) the transverse component of the radiation pressure force
exceed that of an ideal flat reflective sail when the wavelength-averaged
deviation angle exceeds $50.3^\circ$.
Future work in this area may include an optimization analysis of sailcraft acceleration 
that includes internal and external reflections, material absorption and dispersion, 
polarization, spatial coherence, and particularly, alternative beam deviation mechanisms such as 
reflective or transmissive metasurfaces 
\cite{20Hasman2002,21Capasso2011,22Grbic2013,23Capasso2015,24Menon2016,25Bozhevolnyi2018,25Caloz2019,26Magnusson2020,27Xing2021}
and highly birefringent thin geometric phase films such as cycloidal diffractive waveplates
\cite{28Tabiryan2009,29Tabiryan2017,30Tabiryan2021}.
Optimization approaches must include both the momentum transfer efficiency
of the sail, but also the impact on the total mass of the sailcraft.  
For example, advanced materials may afford added functionality that
allows space flight hardware such as heat radiators, photovoltaic, antennae, 
or attitude control devices to be replaced with lower mass elements that
are integrated into the sail.

\section{Acknowledgements} 
This research was supported by NASA Innovative Advanced Concepts Program (NIAC), Grants 80NSSC18K0867 and 80NSSC19K0975.  The author is grateful to the following scientists for useful discussions related to this work:
This report benefitted from discussions with
Les Johnson and Andy Heaton (NASA Marshall Space Flight Center, Huntsville, AL),
Nelson Tabiryan (BEAM Co., Orlando, FL),
Ying-Ju Lucy Chu, Amber Dubill, and Prateek Srivastava (Rochester Institute of Technology, Rochester, NY),
Seongsin Margaret Kim and Anirban Swakshar (University Alabama, Tuscaloosa, AL), 
and Rajesh Menon (University of Utah, Salt Lake City, UT)



\begin{thebibliography}{1}
 \newcommand{\enquote}[1]{``#1''}
 
\bibitem{1Schagrin1974}
   M. L. Schagrin , 
   \enquote{Early observations and calculations on light pressure},  
   American Journal of Physics \textbf{42}, 927-940 (1974).
   
\bibitem{2PlanetarySociety2021}   
	\enquote{The Story of LightSail, Part 1},
	The Planetary Society,
	Accessed 14 Nov 2021,
	\url{https://www.planetary.org/sci-tech/the-story-of-lightsail-part-1}.
	
\bibitem{3Prialnik2009}
 	D. Prialnik,
	\emph{An Introduction to the Theory of Stellar Structure and Evolution},
	(Cambridge University Press, 2009).
   
\bibitem{4Tsiolkovsky1921}
	K. E. Tsiolkovsky,
	\enquote{Extension of man into outer space} (1921)
	[Also in Proc. Symp. Jet Propulsion \textbf{2},
	United Scientific and Technical Presses (1936)].
	
\bibitem{5Tsander1924}
	K. Tsander,
	\enquote{From a scientific heritage },
	Aviation Week \& Space Technology, \textbf{145}, 44-46 (1924).
	
\bibitem{6McInnes2004}
	C. R. McInnes,
	\emph{Solar Sailing: Technology, Dynamics and Mission Applications}
	(Springer, 2004).
	
\bibitem{7Macdonald2006}
	M. Macdonald, G. Hughes, C. McInnes, A. Lyngvi, P. Falkner, A. Atzei, 
	\enquote{Solar polar orbiter: A solar sailing technology reference study}, 
	J. Spacecr. Rockets \textbf{43}, 960-972 (2006).
	
\bibitem{7Macdonald2014}
	M. Macdonald, 
	\emph{Advances in Solar Sailing},
	(Springer Verlag 2014)

\bibitem{8Macdonald2007}
	M. Macdonald, C. R. McInnes, and B. Dachwald,
	\enquote{Heliocentric solar sail orbit transfers with locally optimal control laws}, 
	J. Spacecr. Rockets \textbf{44}, 273-276 (2007).

\bibitem{9Dubill2020}
	A. Dubill, 
	\enquote{Attitude control for circumnavigating the sun with diffractive solar sails},
	(Thesis), Rochester Institute of Technology.
	ProQuest Dissertations Publishing, 27961443, (2020).
		
\bibitem{9Swartzlander2021}
	A. L. Dubill and G. A. Swartzlander, Jr.	
	\enquote{Circumnavigating the sun with diffractive solar sails},
	Acta Astronautica \textbf{187}, 190-195 (2021).

\bibitem{10Johnson2014}	
	L. Johnson, G.A. Swartzlander, Jr., A.B. Artusio-Glimpse, 
	\enquote{An Overview of Solar Sail Propulsion within NASA},
	in Advances in Solar Sailing, 15-23
	M. Macdonald, editor
	(Springer Praxis, 2014).
	
\bibitem{11Spencer2020}
	D. Spencer, B. Betts, J. Bellardo, A. Diaz, B. Plante, J. Mansell,
	\enquote{The lightsail 2 solar sailing technology demonstration},
	Adv. Space Res. \textbf{67}, 2878-2889 (2020).
	
\bibitem{12Mengali2018}	
	M. Bassetto, L. Niccolai, A. A. Quarta, and G. Mengali,
	\enquote{Logarithmic spiral trajectories generated by solar sails},
	Celest Mech Dyn Astr \textbf{130}, 1-24 (2018). 
	
\bibitem{12Swartzlander2017}
	G. A. Swartzlander, Jr.,
	\enquote{Radiation pressure on a diffractive sailcraft},
	 J. Opt. Soc. Am. B  \textbf{34}, C25-C30 (2017).
	 
\bibitem{13Sauer2000}
	C.J. Sauer, 
	\enquote{Solar sail trajectories for solar polar and interstellar probe missions},
	Adv. Astronaut. Sci. \textbf{103}, 547–562 (2000).
	
\bibitem{14Vulpetti2015}emph
	G. Vulpetti, L. Johnson, G. L. Matloff, 
	\emph{Solar Sails: A Novel Approach To Interplanetary Travel}
	(Springer, 2015)
	
\bibitem{15Friedman2015}	
	Louis Friedman,
	\emph{Human Spaceflight, From Mars to the Stars}
	(University of Arizona Press, 2015).	
	
\bibitem{16Tsuda2013}
	Y. Tsuda, O. Mori, R. Funase, H. Sawada, T. Yamamoto, T. Saiki, T. Endo, K. Yonekura, H. Hoshino, J. Kawaguchi,
	\enquote{Achievement of IKAROS — Japanese deep space solar sail demonstration mission},
	Acta Astronautica \textbf{82}, 183-188 (2013).
	
\bibitem{17Johnson2011}	
	L. Johnson, M. Whorton, A. Heaton, R. Pinson, G. Laue, C. Adams,
	\enquote{Nanosail-d: A solar sail demonstration mission},
	Acta Astronautica \textbf{68}, 571-575 (2011).
	
\bibitem{18Lightsail2}
	\enquote{Lightsail 2 Mission Control},
	\url{https://secure.planetary.org/site/SPageNavigator/mission_control.html}
	(Accessed 9 June 2022).
	
	
\bibitem{19Chu2018}
	Y.-J.L. Chu, E. Jansson, G.A. Swartzlander, 
	\enquote{Measurements of radiation pressure owing to the grating momentum}, 
	Phys. Rev. Lett. \textbf{121}, 063903 (2018).

\bibitem{19Chu2019}
	Y.-J.L. Chu, N.V. Tabiryan, G.A. Swartzlander, 
	\enquote{Experimental verification of a bigrating beam rider}, 
	Phys. Rev. Lett. \textbf{123}, 244302 (2019).

\bibitem{19Srivastava2019}
	P.R. Srivastava, Y.-J.L. Chu, G.A. Swartzlander, Jr.
	\enquote{Stable diffractive beam rider},
	Opt. Lett. \textbf{44}, 3082-3085 (2019).
	
\bibitem{19Atwater2019}
	O. Ilic, H. A. Atwater,
	\enquote{Self-stabilizing photonic levitation and propulsion of nanostructured macroscopic objects},
	Nat. Photonics \textbf{13}, 289–295 (2019). 

\bibitem{19Srivastava2020}
	P.R. Srivastava, G.A. Swartzlander, Jr.
	\enquote{Optomechanics of a stable diffractive axicon light sail},
	Eur. Phys. J. Plus \textbf{135},  570 (2020).

\bibitem{19Davoyan2021}
	A.R. Davoyan, J.N. Munday, N. Tabiryan, G.A. Swartzlander, Jr., L. Johnson
	\enquote{Photonic materials for interstellar solar sailing},
	Optica \textbf{8}, 722-734 (2021).

\bibitem{19Swartzlander2022}
	G.A .Swartzlander, Jr.,
	\enquote{Diffractive solar sails},
	J. Physics: Photonics (To appear 2022).
		
\bibitem{19Maxwell1873}
	J. C. Maxwell, 
	\emph{A Treatise on Electricity and Magnetism, Vol. 2} 
	(Macmillan and Co., 1873).
		
\bibitem{20Hasman2002}
	Z. Bomzon, G. Biener, V. Kleiner, and E. Hasman,
	\enquote{Space-variant Pancharatnam–Berry phase optical elements with computer-generated subwavelength gratings},
	Opt. Lett. \textbf{27}, 1141-1143 (2002).
	
\bibitem{21Capasso2011}
	N. Yu, P. Genevet, M. A. Kats, F. Aieta, J.P. Tetienne, F. Capasso and Z. Gaburro,
	\enquote{Light Propagation with Phase Discontinuities: Generalized Laws of Reflection and Refraction},
	Science \textbf{334}, 333-337 (2011).
	
\bibitem{22Grbic2013}
	C. Pfeiffer and A. Grbic
	\enquote{Metamaterial Huygens’ Surfaces: Tailoring Wave Fronts with Reflectionless Sheets},
	Phys. Rev. Lett. \textbf{110}, 197401 (2013).
	
\bibitem{23Capasso2015}
	F. Aieta, M. A. Kats, P. Genevet, and F. Capasso,
	\enquote{Multiwavelength achromatic metasurfaces by dispersive phase compensation}
	Science \textbf{347}, 1342-1345 (2015).
	
\bibitem{24Menon2016}
	P. Wang, N. Mohammad, and R. Menon, 
	\enquote{Chromatic-aberration-corrected diffractive lenses for ultra-broadband focusing}
	Scientific Reports \textbf{6}, 21545 (2016).
	
\bibitem{25Bozhevolnyi2018}
	F. Ding, A. Pors and S. I Bozhevolnyi
	\enquote{Gradient metasurfaces: a review of fundamentals and applications},
	Rep. Prog. Phys. \textbf{81}, 026401 (2018).
	
\bibitem{25Caloz2019}
	K. Achouri, O. V. Céspedes, C. Caloz, 
	\enquote{Solar “Meta-Sails” for Agile Optical Force Control}," 
	IEEE Transactions on Antennas and Propagation, \textbf{67}, 6924-6934 (2019).
	
\bibitem{26Magnusson2020}
	G. Xing, S. Zhang, and R. Magnusson,
	\enquote{Leaky Bloch modal evolution of wideband reflectors with zero-contrast 
	gratings from symmetric trapezoid to triangle ridge shapes},
	Optical Engineering \textbf{59}, 127102, 1-10 (2020).

\bibitem{27Xing2021}
	G. Xing, S. Zhang, X. Mi, and R. Zhang, 
	\enquote{Design and analysis of broadband guided-mode resonant reflectors with coated 
	triangular and trapezoidal profiles in TE polarization}, 
	Opt. Express 29, 26444-26455 (2021).
	
\bibitem{28Tabiryan2009}
	S. R. Nerisiyan, N. V. Tabiryan, D. M. Steeves, B. R. Kimball, 
	\enquote{Optical axis gratings in liquid crystals and their use for polarization insensitive optical switching},
	J. Nonlinear Opt. Phys. Mater. \textbf{18}, 1–47 (2009).

\bibitem{29Tabiryan2017}
	S. V. Serak, D. E. Roberts, J.-Y. Hwang, S. R. Nersisyan, N. V. Tabiryan, 
	T.J. Bunning, D.M. Steeves, and B.R. Kimball,
	\enquote{Diffractive waveplate arrays},
	J. Opt. Soc. Amer. B \textbf{34}, B56–B63 (2017).
	
\bibitem{30Tabiryan2021}
	N. V. Tabiryan,
	\enquote{Advances in Transparent Planar Optics: Enabling Large Aperture, Ultrathin Lenses},
	Advanced optical materials \textbf{9}, 2001692 (2021)

 \end{thebibliography}
\end{document}